\documentclass[reprint,aps,prl,twocolumn,superscriptaddress,showpacs,floatfix]{revtex4-1}
\usepackage{graphicx,enumerate}
\usepackage{amsmath,epsfig}
\usepackage{amssymb}
\usepackage{amsthm}
\usepackage{color}
\usepackage{hyperref}
\usepackage{breakurl}

\newcommand{\be}{\begin{equation}}
\newcommand{\ee}{\end{equation}}          
\newcommand{\ba}{\begin{eqnarray}}
\newcommand{\ea}{\end{eqnarray}}

\newcommand{\ep}{\varepsilon}

\begin{document}

\vspace*{-4mm}
\title{
{\rm \footnotesize  DESY 17-009,~~DO-TH 17/01} 
\\
The Method of Arbitrarily Large Moments to Calculate Single Scale Processes in Quantum Field Theory}

\author{Johannes Bl\"umlein\footnote{Johannes.Bluemlein@desy.de}}
\affiliation{Deutsches Elektronen-Synchrotron, DESY, Platanenallee 6, D-15738 Zeuthen, Germany}

\author{Carsten Schneider\footnote{cschneid@risc.uni-linz.ac.at}}
\affiliation{Research Institute for Symbolic Computation (RISC), Johannes Kepler University Linz,
Altenbergerstra\ss{}e 69, A-4040 Linz, Austria}

\begin{abstract}
\noindent
We device a new method to calculate a large number of Mellin moments of single scale quantities
using the systems of differential and/or difference equations obtained by integration-by-parts 
identities between the corresponding Feynman integrals of loop corrections to physical quantities.
These scalar quantities have a much simpler mathematical structure than the complete quantity. 
A sufficiently large set of moments may even allow the analytic reconstruction of the whole quantity 
considered, holding in case of first order factorizing systems. In any case, one may derive highly 
precise numerical representations in general using this method, which is otherwise completely analytic.
\end{abstract}

\preprint{   
  DESY 17-009, 
  DO-TH 17/01}

\maketitle

\newcounter{linectr}
\newenvironment{myEnumerate}{\begin{list}{(\arabic{linectr})}{\usecounter{linectr}
\labelwidth1ex\itemsep0ex\labelsep1ex\leftmargin2ex\parskip0.0cm\topskip0cm\partopsep0cm
\listparindent0ex}}{\end{list}}


\def\RR{\mathbb R}
\def\ZZ{\mathbb Z}
\def\NN{\mathbb N}

\def\ep{\varepsilon}

\section{INTRODUCTION}

\vspace*{-3mm}
\noindent
Single scale higher order QED and QCD calculations in the massless \cite{Vermaseren:2005qc,Ruijl:2016pkm} 
and massive \cite{Bierenbaum:2009mv} cases at fixed Mellin moment $n$ are given by polynomials of rational 
numbers and a series of a few special constants, the multiple $\zeta$-values, and possible generalizations 
thereof \cite{MZVX}. This is irrespectively the case, whether or not the functional representation for 
general values of $n$ obeys an equation factorizing in first order, \cite{Vermaseren:2005qc}, or not \cite{ELL}. 
If one 
has access \cite{foot1}
to an algorithm, through which a large number of moments, e.g. $N=2000$ or larger, can be calculated,
the method of guessing, cf.~\cite{GUESSING}, for holonomic problems, which often appear in physics 
applications, allows one to gain one large difference equation describing the 
corresponding problem \cite{Blumlein:2009tj}. If this equation is solvable in difference ring theory 
\cite{Summation} one finds the solution for general values of $n$ for this quantity without any further 
assumptions, e.g. made in \cite{Ruijl:2016pkm,Velizhanin:2014fua}. In any case, significantly more 
moments will allow one to constrain the considered quantity much better numerically using approximation 
methods, e.g. through Chebyshev-polynomials or other interpolation methods.

In the following we describe an algorithm through which the system of differential equations, or 
associated to it, that of difference equations, available by the integration-by-parts relations 
\cite{Chetyrkin:1981qh}, can be used to compute a large number of Mellin moments for the master integrals, 
and through them for the whole problem. 
The 
solution of the associated difference 
equations will need a relatively low number of initial values which have to be provided. 
The corresponding sequences of rational numbers mentioned above actually form the problematic part
in gaining the general $n$ result from the moments, since very involved, and in some cases yet 
unknown, functions span the corresponding sequences. In any case, the method allows either to find 
the one-dimensional distribution from a large but finite amount of moments, or at least to constrain
it numerically at high accuracy. 

We will give a brief illustration of the algorithm in case 
of a system of massive 3-loop master integrals, for which first a larger number of moments is generated,
corresponding difference equations are found and solved for general values of the Mellin variable $n$.  

\section{The Algorithm}
\noindent
Single scale master integrals can be represented as analytic functions $\hat{I}_i(x)=\sum_{n=0}^{\infty} 
I_{i}(n)x^n$ with $1\leq$ $i\leq m$. We aim at computing a large number of coefficients 
$I_{i}(0),I_{i}(1),\dots,I_{i}(s)$.
Usually the coefficients $I_{i}(n)$ depend on the dimensional parameter $\ep$ which itself can be 
expanded in a Laurent series in $\ep$ of a certain order $o\in\ZZ$. We are interested in calculating 
the coefficients $I_{i}^{(k)}(n)\in\RR$ of the expansions
\begin{equation}\label{CENT1}
\hat{I}_i(x)=\sum_{n=0}^{\infty} I_{i}(n)x^n=\sum_{n=0}^{\infty} 
\left(\sum_{k=o}^{\infty}I_{i}^{(k)}(n)\ep^k\right)x^n
\end{equation}
up to a certain degree in $\ep$, $t_i\in\ZZ$. More precisely, we want to compute for $1\leq i\leq m$ 
the initial values
\begin{equation}\label{Equ:InitialValuesS}
I_{i}^{(k)}(0),I_{i}^{(k)}(1),I_{i}^{(k)}(2),\dots,I_{i}^{(k)}(s)\in\RR 
\end{equation}
for the $\ep$-orders $o\leq k\leq t_i$.

In our approach we rely on the property that these unknown functions $\hat{I}_i(x)$ are usually 
described by a coupled system of first oder linear differential equations 
\begin{equation}\label{Equ:DESystem}
D_x \left(\begin{matrix} \hat{I}_1(x)\\ \hat{I}_2(x)\\ \dots\\  \hat{I}_m(x)\end{matrix}\right)
=A\,\left(\begin{matrix} \hat{I}_1(x)\\ \hat{I}_2(x)\\ \dots\\ \hat{I}_m(x)\end{matrix}\right)
+\left(\begin{matrix} \hat{r}_1(x)\\ \hat{r}_2(x)\\ \dots\\ \hat{r}_m(x)\end{matrix}\right)
\end{equation}
with $D_x = d/dx$ and where $A$ is an $m\times m$ matrix with entries consisting of polynomials in $\ep$ and $x$ and 
the 
entries $\hat{r}_i$ can be given in form of the expansions
\begin{equation}\label{Equ:riX}
\hat{r}_{i}(x)=\sum_{n=0}^{\infty} r_{i}(n)x^n=\sum_{n=0}^{\infty} \left(\sum_{k=o}^{\infty}
r_{i}^{(k)}(n)\ep^k\right)x^n
\end{equation}
with $r_{i}^{(k)}(n)\in\RR$ for $n\in\NN$. Another important assumption is that the coefficients 
$r_{i}^{(k)}(n)$ can be determined efficiently by using either the method under consideration in a 
recursive fashion or by using, e.g., symbolic summation and integration techniques~\cite{Summation,VLadders}. 
Further, we assume that for reasonable small numbers $s'_i$ the coefficients $I_{i}^{(k)}(n)$ with 
$0\leq n\leq s'_i$ can be computed up to certain $\ep$-degrees $t'_i$, i.e., $o\leq k\leq t'_i$ as a 
preprocessing step. Given this input we introduce the following efficient algorithm that computes the 
coefficients~\eqref{Equ:InitialValuesS} up to large values of $n$.

\begin{enumerate}
\item[(1)] Using decoupling algorithms~\cite{UNCOUPL,OreSys} transform the system~\eqref{Equ:DESystem} 
symbolically to one scalar linear differential equation 
 \begin{equation}\label{Equ:ScalarDE}
\sum_{k=0}^m
 a_{k}(x,\ep) D^{(k)}_x \hat{I}_1(x)
 =\sum_{i=1}^m d_i(x,\ep) \hat{r}_i(x),
 \end{equation}
where the $a_{i}(x,\ep)$ and $r_i(x)$ are polynomials in $x$ and $\ep$; in addition, one obtains identities 
of the form  
 \begin{eqnarray}\label{Equ:RemainingInt}
 \hat{I}_i(x)&=&\sum_{j=2}^m\sum_{k} e_{i,j,k}(x,\ep)D^k_x \hat{I}_j(x)
\nonumber\\ &&
+\sum_{j=1}^m\sum_{k} f_{i,j,k}(x,\ep)D^k_x \hat{r}_j(x)
 \end{eqnarray}
 for $2\leq i\leq n$ where the $e_{i,j,k}(x,\ep)$ and $f_{i,j,k}(x,\ep)$ are rational functions in $x$ and 
$\ep$.
\end{enumerate}
\noindent 
The algorithm proceeds now as follows: Compute in steps~(2)-(4) the initial values for $\hat{I}_1(x)$ 
by using the scalar differential equation~\eqref{Equ:ScalarDE}, and compute afterwards in step~(5) the 
initial values for the remaining integrals with the given formulas~\eqref{Equ:RemainingInt}.

\begin{enumerate}
 
 \item[(2)] Plug $\hat{I}_1(x)=\sum_{n=0}^{\infty}I_1(n)x^n$ and $\hat{r}_i(x)=\sum_{n=0}^{\infty}r_i(n)x^n$ 
into~\eqref{Equ:ScalarDE} and eliminate $D_x$ by using the property $D_x \sum_{n=0}^{\infty} h(n)x^n
=\sum_{n=1}^{\infty} n\,h(n)x^{n-1}$ for a power series $\sum_{n=0}^{\infty} h(n)x^n$. Then by coefficient 
comparison w.r.t.\ $x^n$ and using appropriate shifts one gets a linear recurrence of the form
 \begin{equation}\label{Equ:ScalarREC}
 \sum_{k=0}^d 
b_{k}(n,\ep)I_1(n+k) 
=\rho(n,\ep),
 \end{equation}
with $\rho(n,\ep)=\sum_{j=1}^m\sum_{k=0}^l g_{j,k}(n,\ep)r_j(n+k)$ for some $l\in\NN$ where the 
$b_k(n,\ep)$ 
and $g_{j,k}(n,\ep)$ are polynomials in $n$ and $\ep$. Finally, divide the equation by a factor $\ep^u$ for 
some $u\geq0$ in order to obtain updated polynomials $b_k(n,\ep)$ in $n,\ep$ where not all $b_k(n,0)$ 
with $0\leq k \leq d$ are zero; the $g_{j,k}(n,\ep)$ are now polynomials in $n$ and Laurent polynomials in 
$\ep$. 
 
 \item[(3)] We write the right hand side of~\eqref{Equ:ScalarREC} in the expanded representation 
 \begin{equation}\label{Equ:RhoExpansion}
 \rho(n,\ep)=\sum_{k=o}^{\infty}\rho^{(k)}(n)\ep^k,
 \end{equation}
 with $\rho^{(k)}(n)\in\RR$.
 Since the $r_{i}^{(k)}(n)\in\RR$ in~\eqref{Equ:riX} can be computed efficiently by assumption, the 
coefficients $\rho^{(k)}(n)$ for sufficiently large $k\in\ZZ$ and $n\in\{0,1,\dots,s\}$ can be obtained 
explicitly without any cost. 
 
 \item[(4)] We proceed as follows; compare~\cite{BKSS:12}, Lemma~1. Plugging 
 \begin{equation}\label{Equ:I1Exp}
 I_{1}(n)=\sum_{k=o}^{\infty}I_{1}^{(k)}(n)\ep^k
 \end{equation}
 into~\eqref{Equ:ScalarREC} and doing coefficient comparison
 w.r.t.\ $\ep^{o}$ yield the constraint
 $$ \sum_{k=0}^{d'}
b_{k}(n,0) I_{1}^{(o)}(n+k) 
=\rho^{(o)}(n)$$
for some $d'\leq d$ with $b_{d'}(n)\neq0$. Choose $\delta\in\NN$ such that
$b_{d'}(n)\neq0$ for all $n\in\NN$ 
with $n\geq\delta$. Then we can compute with 
the first values $I_{1}^{(o)}(0),\dots,I_{1}^{(o)}(n+d'+\delta-1)$ 
and the rule
\begin{eqnarray}\label{Equ:RecFormula}
 I_{1}^{(o)}(n)\leftarrow \frac{\displaystyle 
\rho^{(o)}(n')
-\sum_{k=0}^{d'-1} b_{k}(n',0) I_{1}^{(o)}(n'+k)
}{\displaystyle b_{d'}(n',0)}
 \end{eqnarray}
 for $n' = n - d', n\geq d'+\delta$  all the other values in linear time. Now insert~\eqref{Equ:I1Exp} 
with the explicitly 
computed values $I_{1}^{(o)}(n)$ with $0\leq n\leq s$ into~\eqref{Equ:ScalarREC} and move these given values 
to the right hand side. This yields
 $$\sum_{k=0}^{d'}
b_k(n,\ep)I'_1(n+k)
= \sum_{k=o+1}^{\infty}\rho'^{(k)}(n)\ep^k$$
for $I'_1(n)=\sum_{k=o+1}^{\infty}I_{1}^{(k)}(n)$ where the $\rho'^{(k)}(n)$ for $0\leq n\leq s$ and 
sufficiently large $k$ are given explicitly. Now we are in the position to repeat this tactic iteratively to 
compute the remaining coefficients: next $I_1^{o+1}(n)$ with $0\leq n\leq s$, afterwards $I_1^{o+2}(n)$ with 
$0\leq n\leq s$, and eventually $I_1^{t_i}(n)$ with $0\leq n\leq s$.
 
 \item[(5)] Now expand the rational functions $f_{i,j,k}(x,\ep)$ and $g_{i,j,k}(x,\ep)$ 
in~\eqref{Equ:RemainingInt} in a Laurent-series expansion w.r.t.\ $\ep$ and compute their coefficients as 
power series in $x$ up to the necessary orders. Finally, combine all the explicitly given (finite) expansions 
in~\eqref{Equ:RemainingInt} using component-wise addition and Cauchy-products which yield the coefficients 
$I_i^{(k)}(n)$ for $2\leq i\leq m$, $o\leq k\leq t_i$ and $0\leq n\leq s$.
\end{enumerate}

The following remarks are in order; related considerations have been applied also to our algorithms to 
solve coupled systems in terms of nested sums over hypergeometric products~\cite{SymbolicUncouplingMethod1,
VLadders,SymbolicUncouplingMethod2,SymbolicUncouplingMethod3}.
\begin{enumerate}
\item[(i)] 
Our algorithm requires that sufficiently many initial values/coefficients of $\hat{r}_i(x)$ 
and $\hat{I}_1(x)$ up to the right $\ep$-order are computed as a preprocessing step. The necessary bounds 
for these numbers can be determined by analyzing the formulas~\eqref{Equ:ScalarREC} and~\eqref{Equ:RemainingInt} 
accordingly.
\item[(ii)] 
For simplicity, we assumed that only one scalar differential equation arises, as it happens in most 
examples. 
In general, several scalar differential equations might arise, i.e., steps~(2)--(4) have to be executed several 
times.
\item[(iii)] 
One can choose any $\lambda$ with $1\leq \lambda\leq m$ to determine a scalar differential equation in 
$\hat{I}_{\lambda}(x)$. Different choices of $\lambda$ might lead to different recurrences~\eqref{Equ:ScalarREC} 
with different orders $d$ and formulae~\eqref{Equ:RemainingInt} of different size. In our calculations 
we 
analyze for each $\lambda$ ($1\leq\lambda\leq m$) the obtained symbolic formulae and choose that 
$\lambda$ 
for the calculation steps (3)-(5) that serves us best, e.g., to minimize the required $\ep$-orders for 
the 
$\hat{r}_i(x)$ or minimize the the recurrence order $d$.
\item[(iv)] 
Often it is a challenge to determine the first initial values of $I_i^{k}(n)$ to activate the recurrence 
formula~\eqref{Equ:RecFormula}. Thus it is highly desirable to keep the order $d$ in~\eqref{Equ:ScalarREC} 
as small as possible. Since $d$ gets smaller if the degrees of the $a_i(x,\ep)$ w.r.t.\ $x$ 
in~\eqref{Equ:ScalarDE} can be made smaller, the following refinement can be applied. 
Divide~\eqref{Equ:ScalarDE} by the polynomial $h=\gcd_x(a_0,\dots,a_n)$ in $x$. In most applications this 
reduces the degrees of the polynomials $a_i(x,\ep)$ w.r.t.\ $x$ heavily and thus produces recurrences with 
much lower order, e.g., $d=4$ instead of $d=20$ in~\eqref{Equ:ScalarREC}. The price to be paid is to 
determine the coefficients $\rho^{(k)}(n)$ in~\eqref{Equ:RhoExpansion} by extracting the $n$th coefficients 
from the right hand side of~\eqref{Equ:ScalarDE}. Here the more involved calculation steps similarly as 
given in step~(5) must be carried out.
\end{enumerate}

\noindent 
All these aspects have been implemented within our 
package~\texttt{SolveCoupledSystem}~\cite{VLadders,SymbolicUncouplingMethod2,SymbolicUncouplingMethod3} 
which uses subroutines of the summation package~\texttt{Sigma}~\cite{Summation} and the uncoupling package 
\texttt{OreSys}~\cite{OreSys}.

\begin{table}
\begin{tabular}{|c||c||c||c|c||c|c|c|}
\hline 
      & $\varepsilon^{-3}$ 
      & $\varepsilon^{-2}$ 
      & $\varepsilon^{-1}$ & $\zeta_2$  
      & $\varepsilon^{0}$  & $\zeta_2$ & $\zeta_3$ \\
\hline \hline
$J_1$ & 8 & 24 & 48 & 8 & 89 & 24 & 8 \\
\hline
$J_2$ & 8 & 23 & 42 & 8 & 81 & 21 & 8 \\
\hline
$J_3$ & - &  8 & 19 & - & 63 &  - & 5 \\
\hline
\end{tabular}
\caption[]{The number of moments needed to find the associated difference equations.}
\end{table}
\begin{table}
\begin{tabular}{|l||c|c||c|c||c|c|c|c||c|c|c|c|c|c|}
\hline
& \multicolumn{2}{c||}{ $\varepsilon^{-3}$} &
  \multicolumn{2}{c||}{ $\varepsilon^{-2}$} &
  \multicolumn{2}{c|}{ $\varepsilon^{-1}$} & 
  \multicolumn{2}{c||}{ $\zeta_2$} & 
  \multicolumn{2}{c|}{ $\varepsilon^{0}$}  &
  \multicolumn{2}{c|}{ $\zeta_2$}  &
  \multicolumn{2}{c|}{ $\zeta_3$}  \\
\hline
& $O$  
& $R$  
& $O$  
& $R$  
& $O$  
& $R$  
& $O$  
& $R$  
& $O$  
& $R$  
& $O$  
& $R$ 
& $O$  
& $R$ 
\\
\hline 
$J_1$ & 1 & 2 & 2 & 5 & 3 & 11 & 1 & 2 & 4 & 20 & 2 & 5 & 1 & 2 \\
\hline
$J_2$ & 1 & 2 & 2 & 5 & 3 & 10 & 1 & 2 & 4 & 18 & 2 & 5 & 1 & 2 \\
\hline
$J_3$ & - & - & 1 & 2 & 2 &  4 & - & - & 4 & 12 & - & - & 1 & 1 \\
\hline
\end{tabular}
\caption[]{The order and degree of the associated difference equations.}
\end{table}

\section{An Example}

\noindent
Let us consider a set of three 3-loop master integrals $\{\hat{J}_1(x,\ep),\hat{J}_2(x,\ep),
\hat{J}_3(x,\ep)\}$ which obey a $3 \times 3$ inhomogeneous linear coupled system, cf. 
(\ref{Equ:DESystem}), as an illustrative example. They contribute to the massive operator matrix 
element $A_{gg}^{(3)}$ \cite{Ablinger:2014uka}, and are given in~Eqs.~(3.62--3.64).
In a first step, we will use our above method to compute a large series of fixed moments
for $J_{1,2,3}(n,\ep)$, and in a second step we will use this information to guess recurrences for 
these functions. In
our example the integrals $J_i$ have
poles up to $\ep^{-3}$ and we would like to compute the moments up
to the constant term $O(\ep^0)$. Since the number of moments $n_0$
needed to guess a recurrence is not known a priori, the moments are computed up to a reasonable 
number. If this is not
sufficient, further moments can be computed reusing the previous calculations.
E.g., using our algorithm the time to generate
$n_0=500$ moments amounts to 48 sec and $n_0=2000$ moments
require 569 sec in the present example. In a nutshell, higher moments
can therefore easily be produced. 

The guessing method to find the minimal difference equation requires the numbers $n_0$, depending on 
$\varepsilon$ and whether the respective set corresponds to the purely rational term or a respective 
contribution $\propto \zeta_i$. The values are summarized in Table~I. Here the highest number turns out to be 
$n_0 = 89$. The guessing method \cite{GUESSING} yields a difference equation
\begin{equation}
\sum_{k=0}^O a_k(n) F(n+k) = 0
\end{equation}
for each power in $\varepsilon$. Here $O$ and
$R$ denote the order and degree of the difference equation, i.e. the number of shift-operators and the maximal 
degree of the 
polynomials $a_k(n)$. The orders and degrees of the different difference equations are listed in 
Table~II.
The number of moments always 
includes a sufficiently large safety margin to validate the corresponding difference equations, which may even be 
enlarged. We generate additional moments according to their order to provide the needed initial values to solve 
the difference equations.

In the present case, using the algorithm of Ref.~\cite{VLadders}, all difference equations can be solved
completely using difference ring theory  \cite{Summation}. Besides rational terms in $n$ the 
result for 
$J_1(n)$ and $J_2(n)$ is spanned by the harmonic sums \cite{HSUM} up to $S_{1,2}(n)$. $J_3(n)$, furthermore, 
contains the finite binomial sum
\begin{eqnarray}
\frac{1}{4^n} \binom{2n}{n} \left\{\sum_{k=1}^n \frac{4^k S_1(k-1)}{\binom{2k}{k} k^2} - 7 \zeta_3\right\},
\end{eqnarray}
multiplied by a rational term in $n$, see also \cite{Ablinger:2014bra}. All appearing letters building the 
nested
sums forming the master integrals are found automatically. The above example is a simpler one. Let us mention 
that, for comparison, the computational time for the complete pole terms of more involved massive 3-loop 
examples requires the knowledge of about $\sim 1000$ even moments \cite{Blumlein:2009tj} and is estimated to 
amount to several weeks \cite{PREP}.

In case the difference equations would turn out not to be first order factorizable using the algorithm of 
Ref.~\cite{VLadders}, at least the first order factorizable terms would be gained analytically, leaving 
the non-factorizable part behind for further analysis using different methods, like those for elliptic
integrals and their potential generalization.

Having the above number of moments, Eq.~(\ref{CENT1}) yields a first numerical approximation of $\hat{J}_k(x)$
also up to the required power in $\varepsilon$.

The present method is suitable to obtain precise numerical representations in case of various current
massless and massive calculations at 3- and 4-loop order. In case one may solve the associated 
difference equations algebraically, one will even obtain the complete analytic result without any 
prejudice, i.e. a sufficiently large set of scalar moments allows one to fully reconstruct a 
one-dimensional 
distribution. This automatic method is therefore suited to carry out many more higher loop 
calculations in
contemporary elementary particle physics in a completely or at least widely analytic form. 

We would like to thank J.~Ablinger, A.~Behring, A.~De~Freitas, M.~Kauers, and P.~Marquard for discussions.
This work was supported in part by the Austrian Science Fund (FWF) grant SFB F50 (F5009-N15), the European
Commission through contract PITN-GA-2012-316704 ({HIGGSTOOLS}).

\end{document}